\documentclass[lettersize,journal]{IEEEtran}
\usepackage{amsmath,amsfonts}
\usepackage{algorithmic}
\usepackage{array}
\usepackage[caption=false,font=normalsize,labelfont=sf,textfont=sf]{subfig}
\usepackage{textcomp}
\usepackage{stfloats}
\usepackage{url}
\usepackage{verbatim}
\usepackage{graphicx}
\usepackage{booktabs}
\def\BibTeX{{\rm B\kern-.05em{\sc i\kern-.025em b}\kern-.08em
    T\kern-.1667em\lower.7ex\hbox{E}\kern-.125emX}}
\usepackage{balance}
\begin{document}
\title{Perceptual implications of simplifying geometrical acoustics models for Ambisonics-based binaural reverberation}
\author{
    \IEEEauthorblockN{Vincent Martin\IEEEauthorrefmark{1}, Isaac Engel\IEEEauthorrefmark{1}, Lorenzo Picinali\IEEEauthorrefmark{1}}
    \\\IEEEauthorblockA{\IEEEauthorrefmark{1}Dyson School of Design Engineering,  Imperial College London, London SW7 2BD, United Kingdom
    \\}
    
}

\markboth{Preprint}%
{}

\maketitle

\begin{abstract}
Different methods can be employed to render virtual reverberation, often requiring substantial information about the room's geometry and the acoustic characteristics of the surfaces. However, fully comprehensive approaches that account for all aspects of a given environment may be computationally costly and redundant from a perceptual standpoint. For these methods, achieving a trade-off between perceptual authenticity and model's complexity becomes a relevant challenge.

This study investigates this compromise through the use of geometrical acoustics to render Ambisonics-based binaural reverberation. Its precision is determined, among other factors, by its fidelity to the room's geometry and to the acoustic properties of its materials.

The purpose of this study is to investigate the impact of simplifying the room geometry and the frequency resolution of absorption coefficients on the perception of reverberation within a virtual sound scene. Several decimated models based on a single room were perceptually evaluated using the a multi-stimulus comparison method. Additionally, these differences were numerically assessed through the calculation of acoustic parameters of the reverberation. 

According to numerical and perceptual evaluations, lowering the frequency resolution of absorption coefficients can have a significant impact on the perception of reverberation, while a less notable impact was observed when decimating the geometry of the model. 
\end{abstract}


\section{Introduction}
\IEEEPARstart{D}{ifferent} techniques and methods can be employed to simulate the reverberation of a room. In order to categorise them, several factors should be taken into account. A first distinction is between techniques that aim to reproduce a perceptually valid reverberation, and those that seek to reconstruct the exact acoustic sound field from a physical perspective. 

When looking at approaches that seek to reconstruct the exact acoustic sound field, the goal becomes to approximate a solution of the wave equation in the acoustic environment. Computational acoustics methods such as finite-difference time domain (FDTD) \cite{kowalczyk2010room}, finite element method (FEM) \cite{shuku1973analysis}, and boundary element method (BEM) \cite{kirkup2007boundary} are based on the time- and space-discretised solutions of the wave equation. The accuracy of the overall solution is linked to the level of discretisation employed and, thus, to the allocated computational resources.

Another example of an approach that aims at approximating the calculation of real Room Impulse Responses (RIRs) is geometrical acoustics (GA), which is the one employed for this study. It designates the use of several methods, such as ray tracing \cite{krokstad1968calculating} and image-source \cite{allen1979image}, which aim at accurately reproducing the effect of a room given its geometrical description and information about the acoustic characteristics of its materials. 

When rendering RIRs with a GA approach, a model that accounts for every possible feature of a room's geometry and acoustical characteristics is very likely to impact the computational cost of the rendering, requiring a large amount of information about the environment to be simulated. Ultimately, a compromise must be made between the computational requirements, the amount of information allocated to the model, and the perceived accuracy of the auralisation, which refers to the process of recreating an acoustic environment from measured or simulated data \cite{kleiner1993auralization}.

The underlying premise of ray tracing is the assumption that reverberation can be estimated by modeling the frequency-dependent statistical trajectory of energy within a room. This involves envisioning rays originating from a source \cite{savioja2015overview}, interacting with room surfaces, and subsequently being captured by a spatial grid at the receiver's position to depict the directional dispersion of energy characterizing the reverberation within the room. It depict the directional distribution of energy at the receiver position. This characterizes the reverberation within the room for the particular source and receiver configuration. Ray tracing is a high frequency approximation to the solution of the wave equation and in principle more valid for high than for low frequencies.

One of the techniques for reducing the complexity of GA methods is the simplification of the room geometry, where small original surfaces are blended into larger ones. In order to limit the errors in the overall result of the auralisation process, the properties of the decimated surfaces should be accounted for. Also, the volume and equivalent absorption area of the room should be preserved because they affect overall acoustics characteristics. 

Different geometry reduction algorithms have been developed. A complete review of the existing approaches can be found in \cite{siltanen2008geometry}. The methods can be notably divided into two categories: decimation algorithms, which consist of suppressing or merging vertices and/or surfaces; and surface reconstruction algorithms, which aim to reproduce the volumetric topology with a minimal number of surfaces.  The impact of these methods on a ray tracing simulation has been evaluated previously by Abd Jalil et al. \cite{abd2019effect} using numerical metrics such as the Reverberation time (RT) and the Speech Transmission Index (STI). Simplifying the geometrical model by up to an 80\% reduction in the number of surfaces has been found to be potentially perceptually acceptable. One of the goals of this study is to investigate the effects of simplifying room geometry on both the objective metrics of reverberation and on their perceptual impact.

The geometry reduction method used in this study, which is more precisely described in Section \ref{sec:methods}, can be envisioned as a decimation approach, as it consists of manually suppressing small surfaces.



Studies by Brinkmann et al. \cite{brinkmann2019round} and Blau et al. \cite{blau2021toward} have evaluated different room acoustics simulation methods, applying different types of simplifications of the room geometry. A notable example is the evaluation of the RAZR algorithm \cite{wendt2014computationally}, a hybrid approach which combines the image-source method to synthesise early reflections and Feedback Delay Networks (FDNs) for late reverberation, assuming a shoebox-shaped room. Results from these studies demonstrate that, despite the strong simplifications, RAZR still achieves plausible (but not authentic) auralisation results when compared to more detailed simulations, given that the model was fitted to the same frequency-dependent reverberation time. Plausibility refers to the quality of being both convincing and credible, implying that the auralisation possesses characteristics that align with the expected or acceptable auditory attributes of a given environment. It is a less strict criterion as it does not imply an explicit comparison to a reference (e.g., "authenticity") \cite{blau2021toward}. A similar ``shoebox'' assumption is used for Scattering Delay Network (SDN), a method that combines elements of geometric room acoustics with delay networks designed to provide spatialised reverberation corresponding to a given room geometry and wall absorption characteristics \cite{de2015efficient}. SDN can provide perceptually important features of the room response, such as the direct path and first-order reflections, along with a degrading accuracy of higher-order reflections. SDN, when coupled with Head-related transfer function (HRTF) personalisation, was shown to provide levels of ``naturalness'' and ``pleasantness'' that are comparable to full-scale room acoustics simulators \cite{djordjevic2020evaluation}. However, similarly to the RAZR algorithm, there is no study which shows that geometry simplification assumption used in this method is authentic from a perceptual point of view.

One of the methods used to encapsulate and simplify the spatial aspects of a reverberant sound field is encoding it in the Ambisonics format \cite{gerzon1985ambisonics,zotter2019ambisonics}. This technique is based on the expansion of a sound field into a truncated series of spherical harmonics, which are a set of special orthogonal functions defined on a spherical surface. The truncation order of the series, also called Ambisonics order, dictates the spatial resolution of the reproduced sound field, as well as the computational requirements of the method. With the accelerated development and adoption of Virtual Reality (VR) technologies in recent years, Ambisonics-based methods have become more and more popular due to computational scalability, efficiency (and simplicity) for head movements and flexibility to adapt to different types of reverberation rendering methods. Notably, Ambisonics has been integrated into VR-focused acoustic simulation engines developed by Oculus \cite{schissler2017efficient} and Google \cite{gorzel2019efficient}. It is then particularly relevant to investigate how computational optimisations and simplifications affect the perceived quality and realism of reverberation, specifically looking at Ambisonics-based rendering.



The present work has a comparable scope to Engel et al.'s study \cite{engel2021perceptual}, which explored the perceptual impact of varying the spatial resolution when rendering Ambisonics-based reverberation in a binaural setting. The present study used a multiple stimulus test with hidden reference and anchor (MUSHRA) protocol to compare the perceptual similarity of 4th and 3rd order Ambisonics-based binaural reverberation. The study was made on two different rooms in the presence of head movements. It demonstrated that there is no perceptual difference between the two orders of Ambisonics-based reverberation, assuming the direct sound was rendered identically for both on a spatially-dense set of head-related impulse responses (HRIRs).

The motivation of Engel et al.'s study and the current study is to improve the understanding of which reverberation characteristics should be computed and reproduced to achieve realistic Ambisonics-based binaural auralisations from a perceptual standpoint. An identical ``hybrid Ambisonics method'' was employed in this study in which the direct sound is rendered through convolution with a spatially-dense set of HRIRs, while the reverberation is rendered in the Ambisonics domain.

In the study by Engel et al. \cite{engel2021perceptual}, the reverberation was rendered through convolution with measured Ambisonics impulse responses, while in the current study \emph{CATT Acoustics} \cite{GAwhitepaper} was used to synthesise reverberation. This software uses image-source and ray tracing methods to generate reverberation from a geometrical model input. Notably, the software is able to output Ambisonic impulse responses, assuming that the geometrical model of the room, as well as the scattering and absorption coefficients of its surfaces and boundaries, are known. A notable challenge associated with this method lies in the need of precise knowledge of the geometrical details and acoustic characteristics of the environment. The focus of the current study is to evaluate the perceptual impact of simplifying the geometry of the model and the frequency content of the absorption coefficients.

While this method and, more generally, GA models are sometimes used for real-time rendering, they are generally computationally demanding and therefore more suitable for offline high-precision acoustics simulations. The reason why they have been chosen for this study is that they allow us to run precise simulations in order to evaluate whether certain aspects of the rendering are perceptually important or not, while keeping other acoustical parameters identical. Once this is understood, it might be possible to exploit these results to inform real-time rendering methods. SDN could be a practical example; its current implementations allow only for shoebox-shaped rooms to be simulated. This study will give relevant insights as to whether shoebox rooms can be considered as a perceptually suitable alternative to more complex geometries.  %







To summarise, the research question addressed by this study is \textit{how much can a model of a given environment be simplified for it to still generate an authentic simulation through GA and Ambisonics-based binaural rendering?}. 
This question will be explored examining two distinct simplifications applied to the input geometrical model of the room.

First, the impact of geometry reduction on the auralisations will be explored, not only objectively (quantitatively), but also perceptually, similarly to what was done in a previous study by the same authors \cite{engel2021perceptual}. The works of Siltanen et al. \cite{siltanen2008geometry} and Abd Jalil et al. \cite{abd2019effect} have already focused on evaluating geometrically-reduced (GR) models with objective criteria such as the Speech Transmission Index, Lateral Energy Fraction, and Early Decay Time. In this study, the evaluation is focused on perceptual similarity through a MUSHRA paradigm \cite{bs20151534}. It is expected that reducing the number of surfaces will impact the spatial and spectral content of the rendered reverberation, both numerically and perceptually, but the relation between model decimation and perceptual outcomes is still very much unknown.


Second, the impact of reducing the frequency resolution of the acoustics parameters of the model is investigated. This addresses another potential limitation of the GA approach, which is the need for accurate information about the absorption coefficients of the materials within the space. These are usually reported for eight separate frequency bands, but such data is rarely available and is rather difficult to extract, especially in real-time (as may be needed in an AR scenario). Simulations could be simplified by using fewer (broader) frequency bands. The current study evaluates whether incrementally decreasing the number of frequency bands, down to a single broadband frequency-averaged value, has a numerical impact on the output, as well as a perceptual effect. It is expected that such a manipulation would modify the spectral content of the reverberation. More precisely, it is expected that it would even out the frequency content of the reverberation (i.e., minimise differences between bands). The same numerical analyses and perceptual evaluations used for the GR models were used to evaluate these models.



The manuscript is organised as follows; a first section presents the experimental paradigm of the study, how auditory stimuli were rendered through a GA approach, and how the decimation of input information and computational needs were integrated. A numerical evaluation of the differences in terms of spectral and spatial features between the generated stimuli is then presented, followed by a description of the design of the perceptual experiment. Finally, the results of the perceptual evaluations are presented and discussed.

\section{\label{sec:methods} Methods}

This section introduces the chosen evaluation paradigm, providing details about the acoustic environment simulations through GA, as well as the choice of the anechoic stimuli. The process of modifying the model for the different conditions and generating RIRs for various source-listener positions is described in detail, as well as the conversion of synthetic RIRs into binaural stimuli. Finally, a brief overview of the experimental setup and conditions is given.

\subsection{Perceptual evaluation paradigm}

A double-blind listening test paradigm was employed, based on the MUSHRA method \cite{bs20151534}. The reference stimulus was produced with a ``reference'' model, and an anchor stimulus was produced with an ``anchor'' model, both described in Subsection \ref{subsec:rirgen}. Participants were asked to rate the similarity of each stimulus to a reference on a scale with a unique rating ranging from 0 to 100, where the latter meant ``identical to the reference'' .




Two different categories of models were evaluated. For the first category, a geometry reduction (GR) was applied by removing increasingly large objects and surfaces. For the second category, a band reduction (BR) of the number of frequency bands used for the absorption coefficients of the various included surfaces was reduced. How these models were developed is presented in Section \ref{subsec:rirgen}. 
For the evaluation, three different receiver positions were employed, and two different audio stimuli (see Section \ref{subsec:Audio}), each evaluated independently in separate trials. A total of 12 trials were performed by each participant, one for each combination of stimulus type (speech, music), listener position (1, 2, 3), and model category (BR, GR). Therefore, BR and GR models were not compared directly within the same trial.

\subsection{Participants}

Twenty participants (13 males, 7 females) were recruited. Ten had little to no experience with spatial audio, while the others were highly experienced. All of them declared having no hearing impairments, and were asked to complete a questionnaire in which they reported their gender, age, and previous experience with audio, providing informed written consent before taking part in the experiment. No audiometric test was conducted to supplement the participants' self-declaration of having no hearing impairments.

The study and methods followed the tenets of the Declaration of Helsinki. Approval for the experiment design was given by the Science, Engineering, and Technology Research Ethics Committee (SETREC, reference: 21IC6923) at Imperial College London. Data acquisition followed the general data protection regulation.

\subsection{\label{subsec:Audio} Audio Material \& Acoustic Environment}

The same two sets of anechoic recordings used by Engel et al. \cite{engel2021perceptual} were employed in this study, in order to ensure consistency among these research efforts and produce more comparable results.

\begin{figure}[t]
\includegraphics[width=2in]{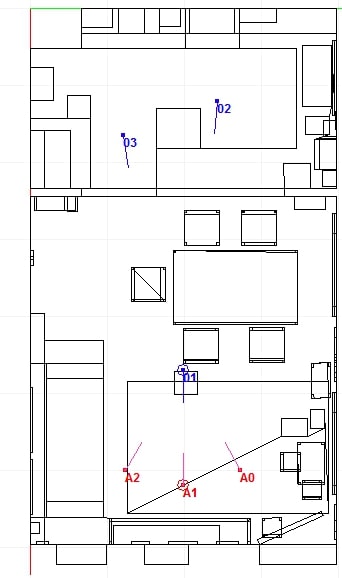}
\centering
\caption{\label{fig:floorplan}{Floor plan of the room used for the experiment. The room consists of an open-plan kitchen, and a living room space separated by a kitchen counter. The blue dots (1-3) show the three different listener positions. The positions of the sound sources are represented by the red dots (A0-2). The A0 position is assigned to the speech recording. The piano recording is emitted from the A0 position for the music stimuli, the drums from A1, and the saxophone from A2.}}
\end{figure}

\begin{itemize}
    \item A music recording of a performance of ``Take Five'' by Desmond \cite{desmond1961take}, consisting of three dry recordings: piano, drum kit, and saxophone. Each source was rendered in a different position (see Figure \ref{fig:floorplan}). This 3-channel recording has been cropped to a length of 7 seconds.
    \item A speech recording from the Music for Archimedes collection \cite{hansen1991making} of a single female speaker, cropped at a length of 5 seconds.
\end{itemize}

The choice of the environment simulated for the experiment was dictated by the need to have a real room with relatively varied acoustic characteristics. A living room with an open kitchen space was chosen (Volume $V=71$m$^3$, measured reverberation time $RT60(1kHz)=0.5$s), which included elements such as wooden floors and carpeted areas, tiled and plasterboard walls, two separate ceiling sections, a kitchen area with hard and reflective surfaces, and a living room area with absorbing surfaces and elements (e.g., bookshelves, sofa, etc.). A single source position was used for the speech stimuli, while three different positions were used for the music sources. Three listener positions were also selected in different areas of the space, close to surfaces with different acoustic characteristics, and at different distances from the sources (see Figure \ref{fig:floorplan}).


\subsection{\label{subsec:rirgen} Models \& RIRs generation} 

\begin{figure}[ht]
\includegraphics[width=\columnwidth]{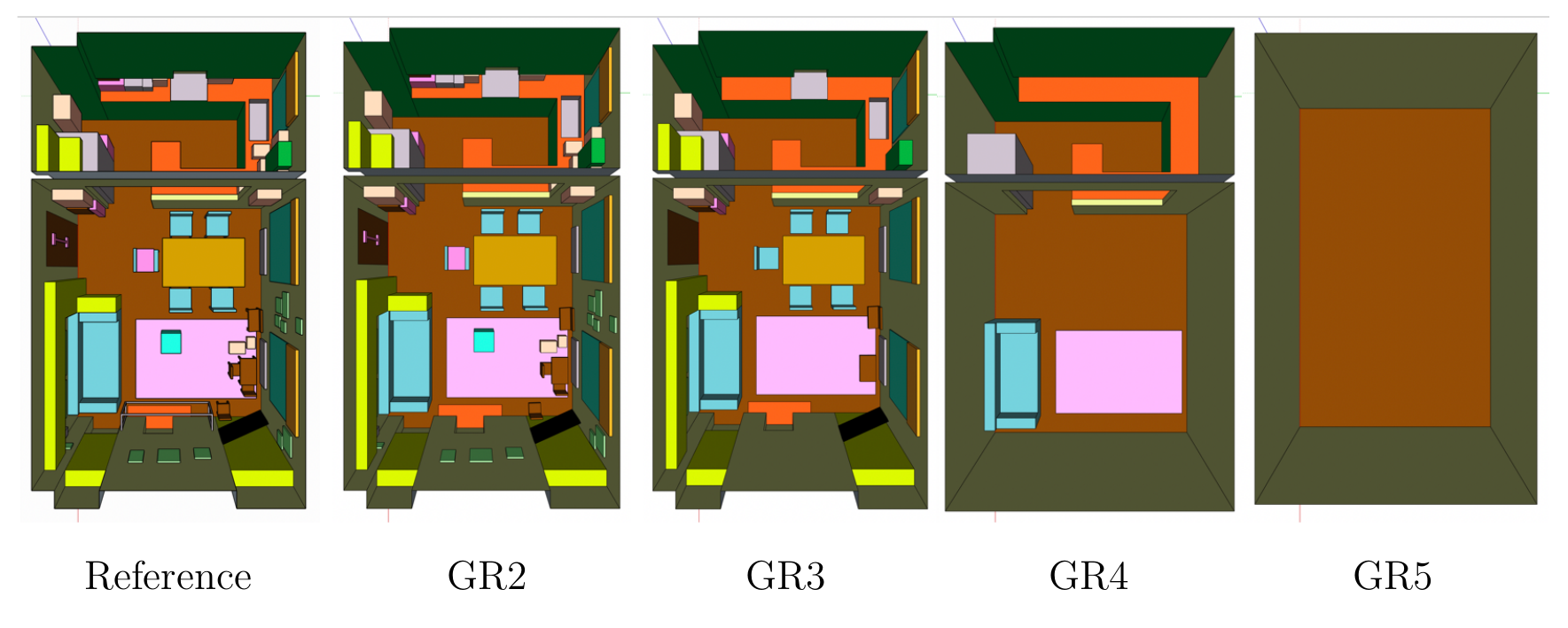}
\caption{\label{fig:GRmodels}{Overhead views of the different GR models from the reference (left), to GR5 ``shoebox'' model (right).}}
\end{figure}

To investigate the impact of geometry reduction, five different GR models of the room described above were prepared using the computer-aided design (CAD) software \textit{SketchUp}, and are illustrated in Figure \ref{fig:GRmodels}. A ``reference'' model was constructed with the highest amount of detail meaningful for GA rendering. Focusing on the absorption and scattering coefficients, standard coefficients for different surfaces were applied. Subsequently, fine-tuning and calibration of these coefficients were undertaken, relying on the measured impulse responses from the actual environment as a benchmark. The objective was to achieve a fitting for the T30 and Early Decay Time decay slopes for each frequency band (see 8 bands in Table \ref{tab:BR-frequency}), with a threshold of error of 1\%. This threshold is below the generally considered just-noticeable differences (JNDs) for these acoustic parameters \cite{iso20093382}.

Four GR models were generated, removing increasingly large objects. For each model, a different threshold was used, which dictated the smallest polygon area allowed, as summarised in Table \ref{tab:GR-Models}. The auralisations were performed with \emph{CATT Acoustics} \cite{dalenback2010characterizing} on a computer with $16$GB of RAM and an 8-core processor running at a $2.6$GHz frequency. The higher the decimation threshold, the fewer polygons in the model, with the extreme case being the shoebox model (GR5), with only six polygons.

The adjustment of the absorption coefficients for the different GR models was a very important operation for the test. In order to avoid significant reverberation time variations across the different models, which would have biased the result as the task would have become an estimation of a perceptual threshold for reverberation time differences, the acoustic absorption coefficients to be employed for the merged planes when reducing the geometrical resolution had to be carefully considered. In order to ensure that the absorption properties of the merged planes accounted for all the neighbouring decimated planes, the coefficients of the GR models were automatically adjusted so that their decay profiles and reverberation times matched those of the reference.




\begingroup

\setlength{\tabcolsep}{10pt} 
\renewcommand{\arraystretch}{1} 

\begin{table}[ht]
\caption{Number of polygons, amount of decimation of the geometry, and time required to compute the auralisation for each GR model.}\label{tab:GR-Models}
\begin{tabular}{llll}
\hline
\textbf{Models} & \textbf{Polygons} & \textbf{Removed surfaces size} & \textbf{Rendering} \\ \hline
Ref.    & 590 & None (Min = 0.01$m^2$ )                       & 249 minutes \\
GR2           & 406 & $<$0.1$m^2$                    & 195 minutes \\
GR3           & 241 & $<$0.4$m^2$                    & 121 minutes \\
GR4           & 66  & $<$0.4$m^2$ and furnitures & 89 minutes  \\
GR5  & 6   & All surfaces                     & 40 minutes  \\ \bottomrule
\end{tabular}

\end{table}
\endgroup



To investigate the impact of reducing frequency resolution, other simplified models were generated. These were based on the previously described reference model, which associated distinct absorption coefficients on eight frequency bands centered around $125$Hz, $250$Hz, $500$Hz, $1$kHz, $2$kHz, $4$kHz, $8$kHz, and $16$kHz. This set of frequency bands is also employed by state-of-the-art simulation methods, such as the RAZR algorithm, in which the FDN parameters are derived to account for a reverberation time defined at the same frequency bands, except for the $125$Hz and $16$kHz bands \cite{kirsch2021spatial}.

Three BR models were developed by reducing the number of frequency bands used for absorption coefficients from 8 bands for the reference, to 4 bands, 2 bands and, finally, a single average value. This process is illustrated in Table \ref{tab:BR-frequency}.

\begingroup

\setlength{\tabcolsep}{10pt} 
\renewcommand{\arraystretch}{1} 

\begin{table*}[ht]
\centering
\caption{Absorption coefficient frequency bands and center frequency for each BR models.}
\label{tab:BR-frequency}
\begin{tabular}{@{}lcccccccc@{}}

\toprule
\textbf{Models}     & \multicolumn{8}{c}{\textbf{Frequency bands' center frequency}}            \\ \midrule
Reference - 8 bands & $125$Hz & $250$Hz & $500$Hz & $1$kHz & $2$kHz & $4$kHz & $8$kHz & $16$kHz \\
BR -  4 bands & \multicolumn{2}{c}{$177$Hz} & \multicolumn{2}{c}{$710$Hz} & \multicolumn{2}{c}{$2840$Hz} & \multicolumn{2}{c}{$11360$Hz} \\
BR - 2 bands        & \multicolumn{4}{c}{$355$Hz}          & \multicolumn{4}{c}{$5680$Hz}       \\
BR - 1 band         & \multicolumn{8}{c}{$1420$Hz}                                              \\ \bottomrule
\end{tabular}

\end{table*}
\endgroup



For all models, synthetic RIRs were generated at the various source-listener pairs via ray tracing, using \emph{CATT Acoustics}. The simulation results were exported as third-order Higher Order Ambisonics (HOA) RIRs. To fulfill the requirements of the MUSHRA protocol \cite{bs20151534}, a set of ``anchor'' Ambisonics RIRs were generated by applying a $2.5\,$kHz low-pass filter to the reference Ambisonics RIRs. Consistently with the other models, only the reverberant part was modified (low-passed), while the direct path signal was not. Removing frequencies above that threshold resulted in a significant decrease of the perceived realism and spatial features of the rendering, allowing the creation of an identifiable (confirmed by the results presented in the following sections, see for example Figure \ref{fig:violinmodelsall}) low-quality anchor which could be used by participants to calibrate the scale of their ratings.  



\subsection{\label{subsec:binrender} Binaural rendering}

The binaural rendering of the auditory stimuli was performed using the \emph{3DTI Toolkit} \cite{cuevas20193d}. It is well known that the spatial resolution requirements are higher for direct sound than for reverberation, while for the direct sound Ambisonics orders between 10 and 30 are found in literature \cite{lubeck2022binaural}, for reverberation orders up to third-order are often deemed perceptually acceptable \cite{engel2021perceptual}. To take advantage of the high spatial resolution of the HRTFs, without being limited by the spatial constraints of the \emph{CATT Acoustics} output (max third order), the direct path component of the auditory stimuli was processed separately from the reverberation. 

To produce the direct path component, the anechoic recordings were convolved with a Head-Related Impulse Response (HRIR) dataset; specifically, the publicly available FABIAN dummy head HRTF \cite{brinkmann2017high} was employed. The reverberant part of the sound was rendered through convolution between the anechoic recordings and third-order Ambisonics RIRs. These had the direct sound component removed by substituting the initial part of the Ambisonics RIRs, up to $4.5$\,ms after the onset, with zeroes, ensuring that the delay between direct and reflected components is preserved at the playback stage. This temporal threshold was calculated analytically using the same approach of Engel et al.'s \cite{engel2021perceptual} minus a safety window of 30 samples (0.68ms). Therefore, a 4.5 ms truncated HRIR was used for the direct sound to avoid overlap with the reverberation. The resulting impulse response contains 198 samples, which exceeds the recommended threshold of 158 samples (128 samples plus an additional 30 samples for inter-aural time delays) for HRIR truncation \cite{sontacchi2002objective}. This threshold was set in order to remove as much energy as possible from the direct sound while keeping as much energy as possible from the early reflections. A portion of the first early reflections might inevitably be excluded due to this cropping method for certain conditions. The estimation of the energy lost in the reverberant part of the generated RIRs was performed by manually determining the time of arrival of the first reflection. In position 1 (see Figure \ref{fig:floorplan}), between $1$\% and $2$\% of the energy of the reverberant part has been removed, between $3$\% and $7$\% in position 2, and between $3$\% and $6$\% in position 3.

In \cite{engel2021perceptual} it is hinted that, at third-order Ambisonics, the quality of the binaural rendering of a room's reverberation is not perceived as different to a fourth order Ambisonics representation, if the direct sound is reproduced with a dense HRIR dataset. Furthermore, it is important to note that no discernible difference was observed through their specific evaluation protocol (MUSHRA), the same protocol utilised in the present study. It is also worth noting that the room examined in the present study falls within the direct-to-reverberant ratio (DRR) range of the rooms examined by Engel et al. \cite{engel2021perceptual} (see Table \ref{tab:DRR} in later section).  Therefore, third-order Ambisonics was employed in this study for the rendering of the reverberation.

The convolution of the reverberant part resulted in an HOA signal, which was then decoded into a binaural format using the so-called ``virtual speakers'' paradigm \cite{zotter2013comparison}. This consists of decoding an HOA signal on an array of virtual loudspeakers surrounding the listener. The minimum number of speakers required for decoding the HOA signal is $(L+1)^2$, where $L$ is the Ambisonics order. In the present case (third-order Ambisonics) an array of $20$ virtual loudspeakers, positioned on the vertices of a dodecahedron, was employed. Based on the position of each virtual loudspeaker on the chosen virtual grid, the associated HRIRs for both ears were identified. Each of them was convolved with that specific loudspeaker feed, and the convolution products were then summed together (separate for the left and right channels), giving the binaural signal for each ear. 

The level of the direct sound and reverberation was determined so that the DRR was kept similar to the real measurements made in the room, for each source-listener pair. The use of the \emph{3DTI Toolkit} \cite{cuevas20193d} allowed for the sound scene to be captured as a binaural recording with a fixed head position. The sound scene was modified dynamically using real-time head tracking information, which resulted in a change of the listener rotation within the simulated environment, and a consequent change of the relative source positions both for the direct sound and for the reverberation (i.e. virtual loudspeakers).

\subsection{Audio playback and VR tools}

\begin{figure}[t]
\includegraphics[width=\columnwidth]{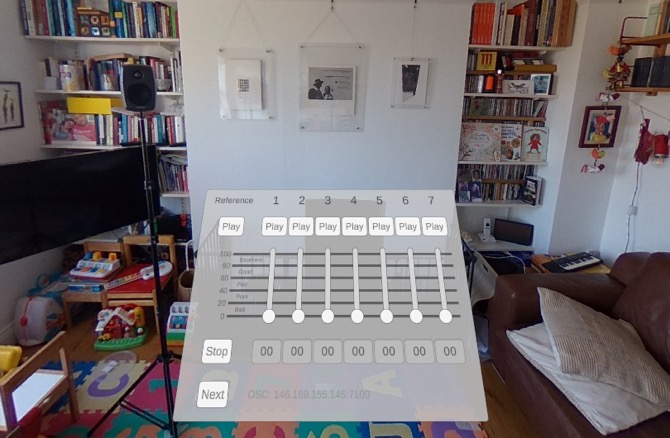}
\caption{\label{fig:interfaceVR}{Interface displayed to participants within the VR headset. The 360° image corresponds to the position of the listener within the environment, and the number of loudspeakers in the picture corresponds to the type of stimuli (1 for speech, 3 for music). This specific image depicts the user view for position 1 and speech signal conditions.}}
\end{figure}

A MUSHRA test user interface (UI), including audio playback and rating controls, was integrated into a simple VR scenario developed with \emph{Unity} and presented via an \emph{Oculus Quest 2} VR headset (see Figure \ref{fig:interfaceVR}). The visuals were generated from 360° pictures taken in the auralized living room, so that participants had a visual cue of the room seen from the correct position and with the right number of visual sources (in the form of loudspeakers - three for music stimuli, one for speech stimuli). Head rotation data provided by the VR headset was wirelessly sent in real time to the \emph{3DTI Tune-In Toolkit}  \cite{cuevas20193d} standalone application, installed on a separate machine, via Open Sound Control (OSC), allowing binaural rendering to account for head movements. 

Sound level was calibrated so that the loudness of the anechoic speech signal spatialised at 1 meter distance was 60dB (LAeq), as recommended by ISO 3382-3 \cite{iso20093382}. Signals were reproduced through a pair of \emph{Sennheiser HD650} headphones connected through a \emph{MOTU Ultralite MK4} interface to the separate audio rendering computer. Headphone equalisation was applied as described by Brinkmann et al. \cite{brinkmann2017high}.

At any time, participants could start the playback of a stimulus linked to a model by clicking on a button within the interface. The sound always began from the start, and was synchronous across the different conditions, so that changing the model did not result in the track restarting from the beginning. The order of the models in the UI was randomised in each trial.

In each of the 12 trials (2 model types $\times$ 2 stimuli types $\times$ 3 listener positions), participants had to evaluate the similarity of the model to the proposed corresponding reference. Trials were presented in a random order. Participants were guided on how to activate each sound using the designated triggers, with the understanding that the reference aligned with the leftmost button. It was specified to participants that they could listen to each sound in any order they preferred and repeat the listening process as many times as desired. Clear instructions were provided regarding the sliders, emphasising their role in assessing similarity with the reference.

\section{\label{sec:4}Numerical analyses and comparisons between the models}

This section illustrates the numerical differences between the auditory stimuli associated with each model, showing how the loss of geometrical information and the number of frequency bands per absorption coefficient affects the spatial and spectral features of the stimuli.
\begin{figure*}[t]
\centering
\includegraphics[width=5in]{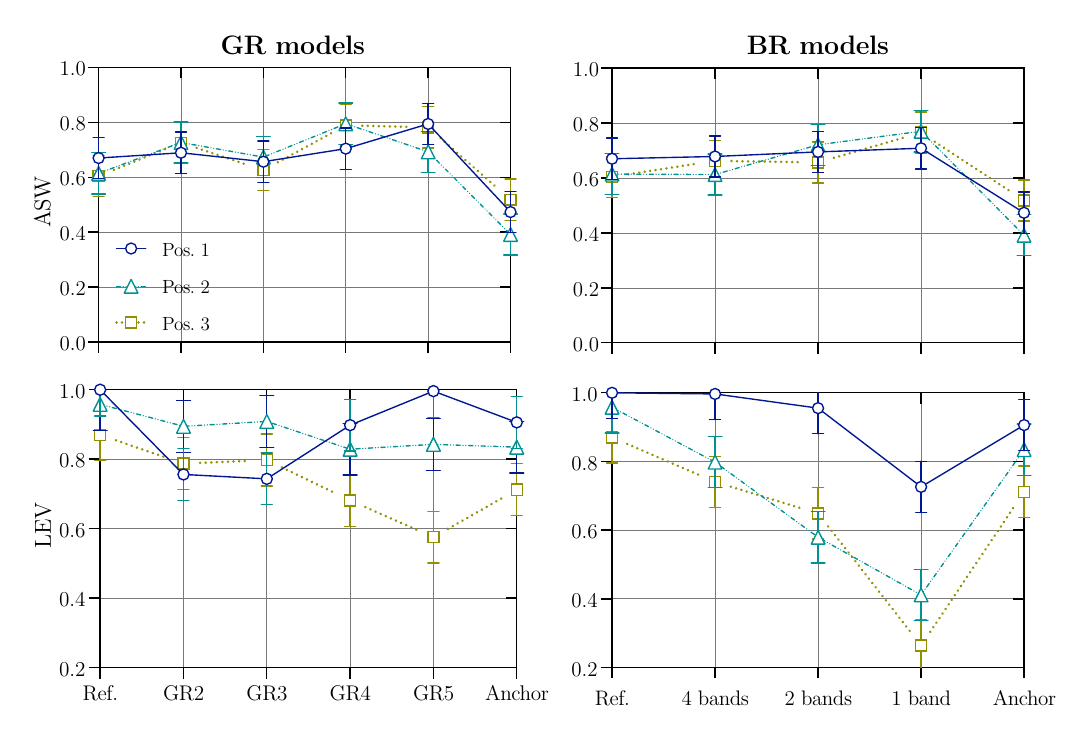}
\caption{\label{fig:ASWLEV}{ASW (upper row) and LEV (lower row) computed for the BRIRs associated to the GR (left) and BR (right) models, for the different listener positions and the source position A0. Bars on each dots represent the just-noticeable difference value associated wtih LEV and ASW}}
\end{figure*}

\subsection{Spatial differences}

To conduct the numerical analysis, synthetic Binaural Room Impulse Responses (BRIRs) were generated for each model by following the binaural rendering procedure described above with a Dirac delta as the input signal. Several perceptually-relevant spatial features associated with these BRIRs were investigated, namely the Apparent Source Width (ASW), the Listener Envelopment (LEV) and the Inter-Aural Cross Correlation (IACC). The first two were calculated following a psychoacoustically motivated prediction model presented in \cite{klockgether2014model}:

\begin{equation}
    ASW = 1 - \frac{\sum _{n,c}IACC^4(n,c)*P(n,c) }{\sum _{n,c}P(n,c)}
\end{equation}

\begin{equation}
    LEV = 1 - \frac{\sum _{n,c}IACC^4(n,c)*(1-P(n,c)) }{\sum _{n,c}P(n,c)}
\end{equation}

With $P(n,c)$ being the weighting function estimating the probability of the presence of a direct sound within the $n$ time frame and the $c$ frequency channel.


Figure \ref{fig:ASWLEV} illustrates the LEV and ASW values obtained for GR and BR models. It is generally considered that the JNDs for both of these parameters are around 0.075 \cite{bradley2011review}. Only differences that could lead to a perceptual impact will be highlighted in respect to this JND value. In this section, significance is related to a difference above the JND value. 

Regarding GR models, for positions 2 and 3 (see Figure \ref{fig:ASWLEV}, left, dotted and dashed lines) no clear patterns emerges in terms of the variation of both parameters. However, it is plausible to suggest that variations in ASW could contribute to perceivable differences between the reference and the GR4 and GR5 models. Similarily, ASW variations could also lead to perceptually relevant differences between the GR2-GR3 models and the GR4-GR5 models. Larger differences can be observed in terms of LEV variations, for example between the reference and GR2 and GR3 models. These also show significantly different LEV values when compared to the GR4-GR5. 

Regarding BR models, for positions 2 and 3 (see Figure \ref{fig:ASWLEV}, right, dotted and dashed lines) the ASW values seem to slightly increase when the number of bands decrease. The anchor exhibits a markedly lower ASW than all other BR models. The reference can be considered here as significantly different from all other models, except for the 4 bands model according to the JND value of ASW. A discernible pattern emerges where the LEV appears to decline alongside the reduction in the number of bands. All models are significantly different from each other according to the LEV JND value.  

For both types of models, the parameters generally exhibit milder variations at position 1, implying that spatial discrepancies are likely to have a smaller impact at this position. These subtle differences can be attributed to the higher Direct-to-Reverberant Ratio (DRR) resulting from the shorter distance between the source and the receiver. In proportion to the consistent direct sound, the reverberation's changes among the models are reduced in this position. Therefore, the parameters are less sensitive to reverberation changes in position 1 if compared with positions 2 and 3.

\subsection{Spectral differences}

\begin{figure*}[!t]
\centering
\includegraphics[width=6in]{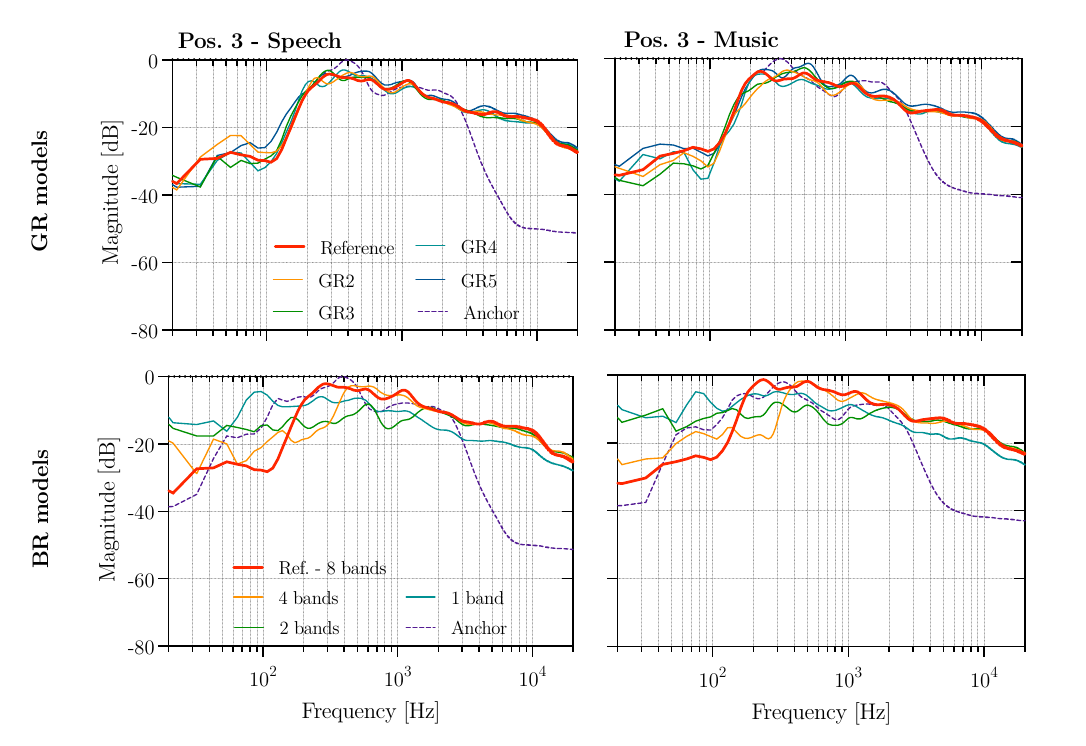}
\caption{\label{fig:lats}{Long-term average spectra of the reverberation tail of the left channel for BRIRs associated with GR models (upper row) and BR models (lower row), for position 3 and for speech (left) and music stimuli (right). The magnitude of each individual diagram has been normalised.}}
\end{figure*}

As mentioned before, the direct sound was processed identically for all models, while the reverberation varied. Therefore, only the spectral differences present in the reverberant part of the BRIRs could induce perceivable differences. Such differences are first illustrated by analyzing the resulting long-term averaged spectra of the reverberation tail for each stimulus (music and speech). Spectra were calculated as the average power spectral density obtained from a series of overlapping 4096-sample Discrete Fourier Transforms (DFTs) after applying 1/3-octave Gaussian smoothing. Figure \ref{fig:lats} depicts the spectrum of the left channels for listener position 3, illustrating the spectral differences between each model's RIRs for both stimulus types. It also highlights the effect of the low-pass filtering processed on the RIRs related to the anchor models.

\begin{figure*}[!t]
\centering
\includegraphics[width=5in]{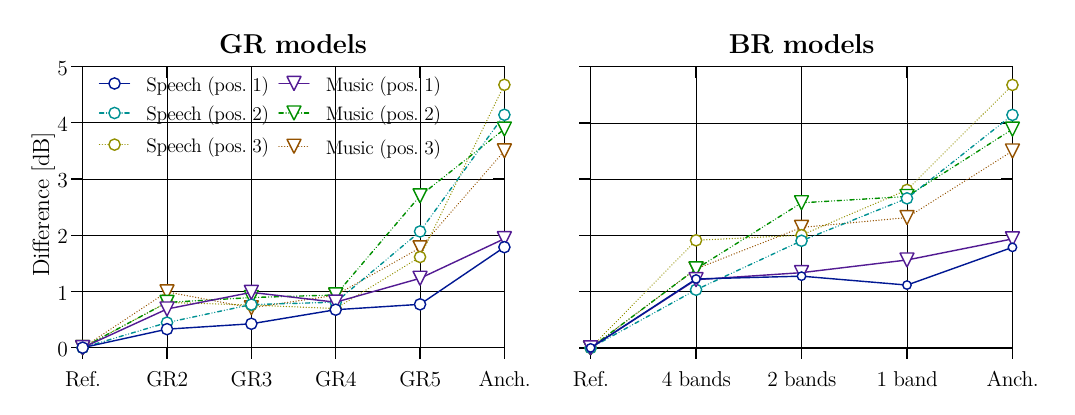}
\caption{\label{fig:spectralscalar}{The absolute difference between each spectrum and the reference, averaged across 42 equivalent rectangular bandwidths (ERBs). GR models are shown on the left and BR models on the right. Results are presented for each of the three positions (indicated in parentheses in the legend) and for the two stimuli (circles for speech, triangles for music).}}
\end{figure*}

A similar analysis was applied to the resulting stimuli after convolution with the synthetic BRIRs. For each condition (stimuli type and listener position), the absolute difference from the reference for each of the 42 Equivalent Rectangular Bandwidths (ERBs) of the resulting long-term averaged spectra of the stimuli was calculated. Figure \ref{fig:spectralscalar} depicts the average of these absolute differences. A similar analysis was used in the study by Engel et al.\cite{engel2022assessing} to assess spectral differences between stimuli.


When decreasing the number of frequency bands per absorption coefficient among BR models, it was expected that spectral features would be affected. Figure \ref{fig:spectralscalar} illustrates that, when the absorption coefficients are averaged to a single value across all frequencies, a maximum difference of 2.8 dB is observed (for listener position 3, speech stimulus).

Surprisingly, the loss of geometrical detail among GR models appeared to have a comparable impact, with a maximum spectral difference between the GR5 model (shoebox) and the reference, comparable to what was observed with the BR models.

For both model types, spectral differences remain relatively small when associated with listener position 1. This is likely caused by the fact that the DRR is higher for this position due to the source being closer to the listener (quantified in Table \ref{tab:DRR}).


\begingroup

\setlength{\tabcolsep}{10pt} 
\renewcommand{\arraystretch}{1} 

\begin{table}[ht]
\centering
\caption{Direct-to-reverberant ratio for each source and receiver position.}
\label{tab:DRR}
\begin{tabular}{llll}
\hline
Receiver position & Position 1 & Position 2 & Position 3 \\ \hline \\
DRR (dB) - A0     & $3.57$     & $-7.57$    & $-8.44$    \\
DRR (dB) - A1     & $3.77$     & $-7.59$    & $-8.23$    \\
DRR (dB) - A2     & $3.81$     & $-7.44$    & $-7.29$    \\ \bottomrule
\end{tabular}

\end{table}
\endgroup

\section{\label{sec:results}Perceptual evaluation results}

First, the results of analyses run with GR models' ratings are presented, followed by the results of BR models. Inferential analysis was performed through a repeated measures analysis of variance (RM-ANOVA). One RM-ANOVA per model type was computed. The ratings of each participant on each stimulus were considered the dependent variable. The RM-ANOVA was conducted with MODEL (6 different GR models or 5 different BR models), STIMULUS (2 types of stimulus) and POSITION (3 listener positions) as within-subject factors. A significance value of $\alpha=0.05$ was used. The effect of the prior experience with spatial audio was studied as a between-subject factor within the RM-ANOVA. It revealed that it had no significant effect on listeners ratings ($F(1,2) = 0.88$ $p = 0.45$) for both types of models.

For each type of model, the results of pairwise comparisons run with Bonferroni t-tests are presented to highlight differences between models in different positions or stimulus types. A Bonferroni correction was applied to the test results. To maintain clarity throughout the section, only a relevant selection (i.e., those showing statistically significant differences) of the pairwise comparisons is reported. The differences between GR models and the reference are highlighted.
 
\subsection{GR models}

\begin{figure}[ht]
\includegraphics[width=\columnwidth]{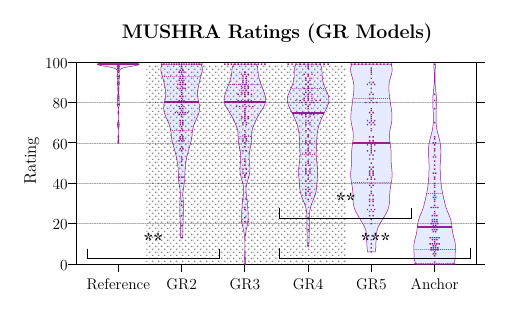}
\caption{\label{fig:violinmodelsall}{MUSHRA ratings of GR models represented by violin plots, showing the probability density of the data and the mean (horizontal line). The ratings across all listener positions and stimulus types are included. The gray area represents a portion of the models whose ratings are not significantly different between each other according to pairwise comparisons performed with t-tests, considering a p-value $> 0.05$. The symbols (*) represent p-values $< 0.05$, (**) represent p-values $< 0.01$, and (***) represent p-values $< 0.001$, indicating the respective levels of statistical significance.}}
\end{figure}

Ratings associated with GR models are shown in Figure \ref{fig:violinmodelsall}. The descriptive statistics illustrated by this figure illustrates that models with a larger amount of geometrical detail obtained higher ratings. The highest rating is consistently obtained with the reference, and the lowest with the anchor. The main independent variable for the RM-ANOVA was the model type, but its interactions with other variables such as listener position, stimulus type, were investigated as well. The results for GR models' ratings are summarised in Table \ref{tab:ANOVA-GR-XP1}.

\begingroup
\setlength{\tabcolsep}{10pt} 
\renewcommand{\arraystretch}{1} 

\begin{table}[ht]
\caption{Results of the RM-ANOVA applied to the ratings of the GR models.} 
\centering
\begin{tabular}{@{}llccl@{}}
\toprule
\textit{\textbf{Effects}} & \textit{\textbf{df}} & \multicolumn{1}{l}{\textit{\textbf{F}}} & \multicolumn{1}{l}{\textit{\textbf{p-value}}} & \textbf{$\eta$$^{2}$} \\ \midrule
\textbf{Geometry reduction}                            & \textbf{5}  & \textbf{93.2} & \textbf{$<$0.001} & \textbf{0.838} \\
\textbf{Position}                         & \textbf{2}  & \textbf{16.5} & \textbf{$<$0.001} & \textbf{0.478} \\
Stimulus type                             & 1           & 0.563         & 0.463              & 0.030          \\
\textbf{Geo. red. * Position}                 & \textbf{10} & \textbf{5.28} & \textbf{$<$0.001} & \textbf{0.227} \\
Geo. red. * Stim. type                     & 5           & 2.122         & 0.135              & 0.105          \\
\textbf{Geo. red. * Stim. type * Pos.} & \textbf{10} & \textbf{2.58} & \textbf{0.006}     & \textbf{0.125} \\ \bottomrule
\end{tabular}

\label{tab:ANOVA-GR-XP1}
\end{table}
\endgroup

Following the MUSHRA recommendations \cite{bs20151534}, initial attention was focused on the sphericity of the ratings associated with each GR model. Therefore, the Huynh-Feldt correction was applied to reduce type I errors as the data did not pass the Mauchly's test of sphericity ($p<0.05$) and showed a Greenhouse-Geisser epsilon higher than $0.75$ ($\varepsilon=0.77$).



\paragraph{Effect of geometry reduction}

The RM-ANOVA found a significant main effect of the geometry reduction. This aligns with the descriptive statistics which shows that highly detailed models are linked to higher MUSHRA ratings (see Figure \ref{fig:violinmodelsall}). In order to find similarity between GR models' ratings, multiple pairwise comparison t-tests were run with a Bonferroni correction for a set of 15 between-ratings associated with each GR model. These results are also illustrated in Figure \ref{fig:violinmodelsall}. These initial pairwise comparisons include ratings from all positions and stimulus types.  Significant differences were found between the reference and all other GR models ($t(15)\geq4.79$, $p_{bonf}<0.001$). Similarly, significant differences were found between the anchor and other GR models ($t(15)\geq11.5$, $p_{bonf}<0.001$) and between the shoebox model (GR5) and all others ($t(15)\geq5.973$, $p_{bonf}<0.001$). These results illustrate that these three models (reference, shoebox and anchor) are judged to be significantly different from all other GR models. Contrarily, no significant differences were found between the GR2, GR3 and GR4 models ($t(15)\leq1.38$, $p_{bonf}>0.05$). These results show that even the smallest geometrical reduction in GR2 has a significant impact on perceptual ratings, but that further reductions in geometry do not result in additional significant differences. Considering the significance of two-way and three-way interactions, these results are further investigated by examining the GR models' ratings in each listener position first, and in each position/stimulus type condition afterwards.

\paragraph{Effect of listener position}

\begin{figure}[ht]
\includegraphics[width=\columnwidth]{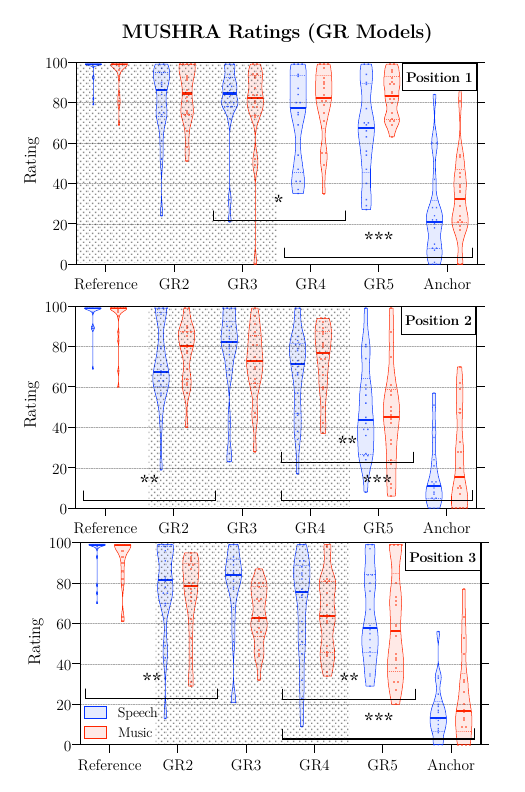}
\caption{\label{fig:violinmodels}{MUSHRA ratings of GR models represented by violin plots, which show the probability density of the data and the mean (horizontal line). Separated per stimulus type (blue: speech, red: music) and for each listener position (top to bottom). The grey area represents a portion of the GR models whose ratings are not significantly different from each other according to pairwise comparisons, with a p-value $> 0.05$. The symbols (*) represent p-values $< 0.05$, (**) represent p-values $< 0.01$, and (***) represent p-values $< 0.001$, indicating the respective levels of statistical significance (grouped across stimulus types).}}
\end{figure}

A significant main effect of position and an interaction effect of geometry reduction and position were found by the RM-ANOVA. Descriptive statistics (see Figure \ref{fig:violinmodels}) hint that in general, at larger source-receiver distances (positions 2 and 3), the different GR models were judged with lower ratings than at a short distance (position 1). Pairwise comparisons t-tests were run between ratings associated with a single model and position to confirm this observation. Considering that no main effect or two-way interaction of the stimulus type were found significant, these pairwise comparisons were run by grouping ratings from both stimulus types. Additional comparisons exploring the effect of the stimulus type in a three-way interaction are presented further.

For each position, the ratings associated with the anchor are significantly different from all other models. For positions 2 and 3, the reference ratings are significantly different from all the other models ($t(153)\geq3.96$, $p_{bonf}\leq0.043$). No significant differences are found between the ratings of the GR2, GR3, GR4 models ($t(153)\leq0.93$, $p_{bonf}>0.05$). In these listener positions, the results confirm that the slightest geometrical reduction in GR2 significantly influenced perceptual ratings, while further geometry reductions did not amplify this difference. Ratings associated with GR5 were found to be significantly different from all the other models ($t(153)\geq5.23$, $p_{bonf}<0.001$). These results confirm that the ``shoebox'' model led to perceptual ratings that differed from all other models, in contrast to the GR2-GR3-GR4 models, which underwent smaller geometrical reduction. For position 1, the reference ratings are not significantly different from GR2 and GR3 ($t(153)\leq2.85$, $p_{bonf}>0.05$). These results confirm that at a short distance the reference is not distinguishable from the first two GR models, but this changes at larger distances, i.e., for positions 2 and 3. 

\paragraph{Effect of stimulus type}

The RM-ANOVA revealed no significant main effects of stimulus type or two-way interaction effects involving stimulus type. However, a significant three-way interaction was found. This suggests that stimulus type can lead to significant differences between models for a given listener position, which were not identified in the previous pairwise comparisons focusing on the main effects and interaction effects of position and model.

To maintain clarity and identify differences or similarities between models' ratings that were not revealed in the previous analyses, only a relevant selection of the pairwise comparisons will be presented.

Within position 1 and the GR5 model, the ratings associated with speech stimuli are significantly lower than those associated with music stimuli ($t(630)\geq-3.77$, $p_{bonf}=0.017$).

Within position 2 and the GR5 model, the ratings associated with speech stimuli are significantly lower than those associated with music stimuli ($t(630)\geq-4.152$, $p_{bonf}=0.025$).

Within position 3 and music stimuli, ratings associated with the GR3 model are significantly lower than those associated with the GR2 model ($t(630)\geq-4.32$, $p_{bonf}=0.001$). This is not observed with ratings associated with speech stimuli, as stated earlier.

These results highlight that despite the significance of the three-way interaction, a clear trend regarding the effect of stimulus type within each model and stimulus type condition does not emerge.

\subsection{BR models}

\begin{figure}[ht]
\includegraphics[width=\columnwidth]{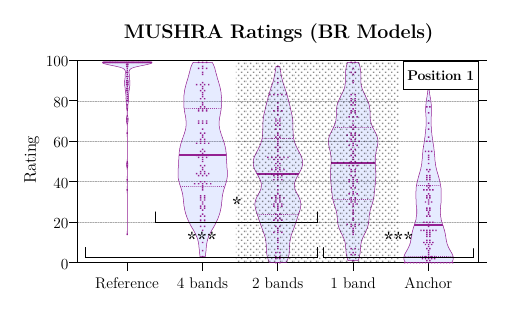}
\caption{\label{fig:violinbandsall}{MUSHRA ratings of BR models represented by violin plots, which show the probability density of the data and the mean (horizontal line). The ratings across all listener positions and stimulus types are included. The gray area represents a portion of the models whose ratings are not significantly different from each other according to pairwise comparisons, with a p-value $> 0.05$. The symbols (*) represent p-values $< 0.05$, (**) represent p-values $< 0.01$, and (***) represent p-values $< 0.001$, indicating the respective levels of statistical significance.}}
\end{figure}

Ratings associated with BR models are shown in Figure \ref{fig:violinbandsall}. The descriptive statistics illustrated by this figure shows that the ratings were higher for models with a large number of frequency bands per absorption coefficient. Moreover, the differences between the reference and the BR models (40 to 60 in similarity rating for BR models, see Figure \ref{fig:violinbandsall})  is larger than the differences found between GR models (60 to 80 in similarity ratings for BR models, see Figure \ref{fig:violinbandsall}). This result seems to show that the band reduction applied here has a larger impact than the geometrical reduction on the perceptual ratings. The results of the RM-ANOVA analysis are summarised in Table \ref{tab:ANOVA-BR-XP1}. Following the MUSHRA recommendation \cite{bs20151534}, the Greenhouse-Geisser correction was applied to reduce type I errors as the data did not pass Mauchly's test of sphericity ($F(1,9) = 0.21$ $p=0.07$) and showed a Greenhouse-Geisser epsilon higher than 0.75 ($\varepsilon=0.76$).

\begingroup

\setlength{\tabcolsep}{10pt} 
\renewcommand{\arraystretch}{1} 

\begin{table}[h]
\caption{Results of the RM-ANOVA applied to the ratings of the BR models.}
\label{tab:ANOVA-BR-XP1}
\begin{tabular}{@{}llccl@{}}
\toprule
\textit{\textbf{Effects}} & \textit{\textbf{df}} & \multicolumn{1}{l}{\textit{\textbf{F}}} & \multicolumn{1}{l}{\textit{\textbf{p-value}}} & \textbf{$\eta$$^{2}$} \\ \midrule
\textbf{Band reduction}                   & \textbf{4} & \textbf{87.7} & \textbf{$<$ 0.001} & \textbf{0.830} \\
\textbf{Position}                & \textbf{2} & \textbf{24.8} & \textbf{$<$ 0.001} & \textbf{0.579} \\
\textbf{Stimulus type}           & \textbf{1} & \textbf{7.22} & \textbf{0.015}     & \textbf{0.286} \\
\textbf{Band red. * Pos.}        & \textbf{8} & \textbf{4.28} & \textbf{$<$ 0.001} & \textbf{0.192} \\
\textbf{Band red. * Stim. type}   & \textbf{4} & \textbf{3.59} & \textbf{$<$ 0.001} & \textbf{0.192} \\
Band red. * Stim. type * Pos. & 8          & 0.81          & 0.596              & 0.043          \\ \bottomrule
\end{tabular}

\end{table}
\endgroup

\paragraph{Effect of band reduction}

The RM-ANOVA shows a significant main effect of the model. It aligns with the descriptive statistics, which illustrates that models with absorption coefficients defined with a larger frequency content are linked to higher MUSHRA ratings (see Figure \ref{fig:violinbandsall}). It can be noted that the difference between BR models and the reference is higher than between GR models and the reference. To investigate further this observation and individual differences between models, inital t-tests were run with a Bonferroni correction for a family of 10 ratings associated with each model (grouping all positions and stimulus types). These results are illustrated in Figure \ref{fig:violinbandsall}. Pairwise comparisons found that most models were different from each other. The only exception is for the ratings of 1-band and 2-band models that are not significantly different from each other ($t(10)\geq1.77$, $p_{bonf}>0.05$). These results are further investigated by examining the ratings associated with each listener position, for each BR model.

\begin{figure}[ht]
\includegraphics[width=\columnwidth]{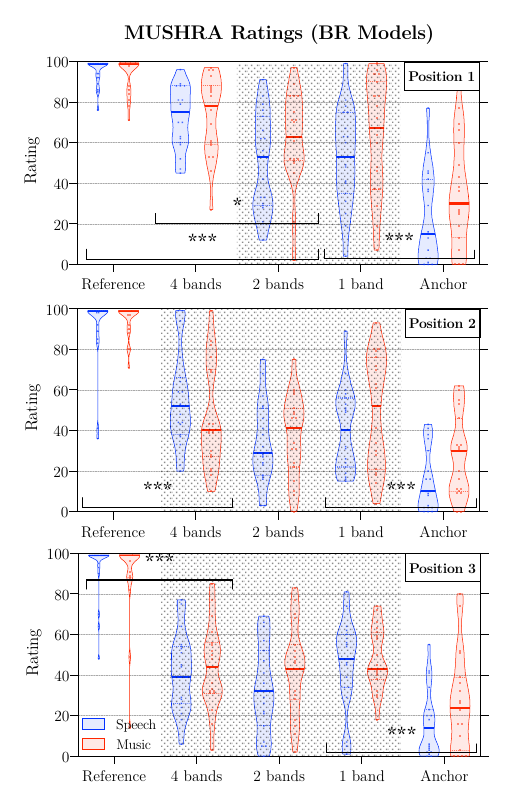}
\caption{\label{fig:violinbands}{MUSHRA ratings of BR models represented by violin plots, which show the probability density of the data and the mean (horizontal line). Separated per stimulus type (blue: speech, red: music) and per listener position (top to bottom). The gray area represents a portion of the BR models whose ratings are not significantly different from each other according to pairwise comparisons, with a p-value $> 0.05$. The symbols (*) represent p-values $< 0.05$, (**) represent p-values $< 0.01$, and (***) represent p-values $< 0.001$, indicating the respective levels of statistical significance (grouped across stimulus types).}}
\end{figure}


\paragraph{Effect of listener position}

A significant main effect of position and an interaction effect of band reduction and position was found by the RM-ANOVA. Similar to what was observed on GR models, at larger source-receiver distances (positions 2 and 3), the different BR models are judged with lower ratings than at a short distance (position 1, see Figure \ref{fig:violinbands}). To investigate individual differences between models and for each position, pairwise comparisons were run between ratings associated with models for a single position. For position 1, 2-band and 1-band models' ratings are not significantly different from each other ($t(10)\geq0.47$, $p_{bonf}>0.05$). For positions 2 and 3, 4-band, 2-band and 1-band model ratings are not significantly different from each other ($t(10)\geq2.96$, $p_{bonf}>0.05$). These results show that reducing the frequency bands from 8 to 4 already generates a strong impact on perceptual ratings. Further reduction in the number of bands does not result in a more significant difference in the ratings.


\paragraph{Effect of stimulus type}

A significant main effect of stimulus type and interaction effect of stimulus type and position were found by the RM-ANOVA.  Pairwise comparisons ran between ratings associated with a single model and stimulus found significant differences for the anchor model, where the music stimuli were judged to have a significantly higher rating than the speech stimuli. Surprisingly, this difference could not be found in the GR model comparison.




\section{Discussion}

The purpose of this study was to investigate if reducing the geometry of a room and the frequency resolution of its absorption coefficients could influence the perception of Ambisonics-based binaural reverberation generated through a GA approach. Several models, created reducing both of these characteristics, were perceptually evaluated. Perceptual results are here discussed in light of the numerical analyses presented in Section \ref{sec:4}.


\subsection{Effect of the geometry reduction}

Two different attributes were numerically evaluated: spatial and spectral features.
Objective spatial metrics (ASW and LEV) illustrated in Figure \ref{fig:ASWLEV} reveal that there are significant differences between GR models for positions 2 and 3 (i.e., at a longer distance from the source than position 1). Notable differences in terms of LEV and ASW (above the just-noticeable differences \cite{bradley2011review}) are present between the Reference and GR2-GR3 models, while GR2-GR3 are different from GR4. This model's parameters are itself different from the shoebox model GR5.


The absolute spectral difference between models and reference illustrated in Figure \ref{fig:spectralscalar} displays a significant difference between the reference and GR2-GR3-GR4 models, themselves different from GR5 and anchor models for positions 2 and 3. Additionally, the GR5 (shoebox) model exhibits spectral differences from the reference similar to those between the 2-band and 1-band BR models. All three models have comparable perceptual ratings. Therefore, it can be inferred that spectral differences are likely the primary cause of the high perceptual differences observed between the reference and the GR5 model, similar to those observed in the BR models.


Results of the perceptual evaluation show that, for Position 1, no significant differences can be found between the reference and the GR2-GR3-GR4 models. This is consistent with the numerical analyses, which show that the variations of spatial and spectral parameters are smaller. At position 1, the DRR is higher (see Table \ref{tab:DRR}), and thus differences caused by the reverberation are less perceptible.

At positions 2 and 3, GR2-GR3-GR4 perceptual results are significantly different from the reference and GR5, but not from each other. Similar results can be found when looking at the results per stimulus type independently. Even removing small surfaces (GR2, $<0.1$m$^2$) seems to be perceptually noticeable. However, GR2 and GR4 ratings are not significantly different, therefore removing all furniture and small surfaces (in GR4, all furniture, and surfaces $<0.4$m$^2$) did not have a more relevant impact on similarity ratings.

This result suggest that, in this context, removing only the smallest surfaces (GR2, 68$\%$ of the reference polygons) is perceptually similar to removing all smaller surfaces and furniture (GR4, 11 $\%$ of the reference polygons). This result is consistent with the differences highlighted by the numerical analyses. Part of these results can be confirmed by the study by Abd Jalil et al. \cite{abd2019effect}, in which the impact of geometry reduction on the auralisation of open-plan office rooms has been quantified through acoustic parameters. Albeit a few differences between the two studies are present (e.g. the use of the image-source method and the fact that they did not process the direct path separately), it was found that removing small surfaces (up to 80$\%$ of the surfaces are removed) has little effect on the chosen metrics (RT60, SPI, and SPL). The conclusions are similar to the ones of the current study; as long as the room's larger surfaces are presented, it is perceptually acceptable to apply significant geometric approximations to the model.

However, differently from this work, the work of Abd Jalil et al. \cite{abd2019effect} showed that removing all furniture had a significant impact on these metrics, notably on the reverberation time (RT) spectral profile. This difference can be explained by considering several factors. Firstly, their simulated rooms are larger than the room used in this study (open-plan offices with a volume of $V>350$m$^3$ vs. $V=71$m$^3$ here) and contain a greater quantity of furniture. Secondly, an additional compensation for the reverberation time was applied here. The frequency content of the absorption coefficients was manually fine-tuned so the spectral profile of RT60 matched the one given by real RIR measurements. This approach was used in order to replicate an individual auralisation process for each GR model, maintaining reverberation and decay profiles consistent with the actual physical environment. It could be inferred that such a strategy could compensate for the perceptual impact of removing a room's furniture. If true, this compensation would result in a significantly simpler approach if compared with having to calculate complex GA models.


The results from this study can be benchmarked looking at other two works where simulated and measured BRIRs were compared \cite{blau2021toward, brinkmann2019round, kirsch2023computationally, starz2022perceptual}. First, It is important to note that one of the main differences between these studies and the current one is the type of perceptual evaluation reports used. In the current study, a direct comparison between different models and a reference stimulus was made in terms of ``similarity''. This could be related to the ``authenticity'' measure used in Brinkmann et al.'s study \cite{brinkmann2019round}, which indicates the presence of any audible difference with a reference. In the study by Brinkmann et al. \cite{brinkmann2019round}, ``plausibility'' ratings were also used. Plausibility refers to the quality of being both convincing and credible, implying that the auralisation possesses characteristics that align with the expected or acceptable auditory attributes of a given environment. It is a less strict criterion as it does not imply an explicit comparison to a reference. It could be related to the evaluation method used in studies by Blau et al. \cite{blau2021toward}, Starz et al. \cite{starz2022perceptual} and Kirsch et al. \cite{kirsch2023computationally}, where participants evaluated stimuli based on different characteristics without any explicit reference comparison.

In the study conducted by Blau et al. \cite{blau2021toward}, participants were tasked with assessing diverse head-tracked simulated BRIRs in contrast to measured BRIRs. Different additional simplifications were explored, including individualised HRTFs and sound source directivity. In the present study, the directionality of the speaker was incorporated into all models, utilising generic HRTFs consistently. 

It is also essential to highlight a key divergence between hybrid methods such as RAZR and the one based on HOA and \emph{CATT Acoustics} here. These hybrid methods utilising FDNs to generate late reverberation by superposing spatially distributed, incoherent sound sources. On the other hand, HOA approaches inherently create an array of coherent, not incoherent, reverberant sources. It is worth noting that despite these differences, both hybrid methods use the image-source method for early reflections. 

Blau et al. \cite{blau2021toward} revealed that methods making assumption of a shoebox regarding the room's geometry did not exhibit a noticeable distinction compared to their reference, which was based on real BRIRs. This echoes similar findings  by Kirsch et al.\cite{kirsch2023computationally} when assessing their method's (based on RAZR algorithm) plausibility against measured BRIRs. This contradicts with the present results that reveal GR5 (shoebox model) to be distinct not only from the reference but also from the other GR models in terms of similarity ratings. Moreover, numerous distinctions in their protocol could potentially account for the reported differences in the output data. Firstly, their evaluation framework diverged from MUSHRA \cite{bs20151534}. In the current study, all models were evaluated within the same trials, assessing their similarity with a reference, on a single scale ranging from excellent to poor. This stands in contrast to the assessment of various characteristics (reverberance, source distance, source direction, source width) from Blau et al's study \cite{blau2021toward}. Additionally, in Blau et al. study, participants were not provided with an explicit reference, resulting in absolute ratings for the characteristics rather than relative ones. Blau et al. also raised concerns about potential biases stemming from their low-quality anchor. It was noted that ratings obtained from the anchor tended to be consistently minimal, possibly explaining the elevated ratings assigned to the shoebox models. The RAZR algorithm was further investigated in a study by 
Starz et al. \cite{starz2022perceptual}. Participants evaluated simulated and real binaural recordings in a scenario where a visual of the auralised room was presented using a head-mounted display. They used a criterion-free method to estimate plausibility via a Yes/No task. This method did not involve a direct comparison between measurements and simulated BRIRs, but it can be considered more strict than evaluating on a scale (as it is in Blau et al.'s study \cite{blau2021toward}). In this study, the authors concluded that the measurements and simulated recordings were indistinguishable in terms of plausibility ratings.

In the current study, the absence of separate evaluation of spatial and spectral attributes prevents from drawing  conclusions about the origin of the observed differences between the shoebox model and the other models. However, as stated earlier, the shoebox model used here (GR5) presents significant spectral differences compared to the reference, which could explain its significantly different perceptual ratings.

These studies offer different conclusions from a round-robin comparison performed by Brinkmann et al. \cite{brinkmann2019round} in which four different algorithms were used to generate BRIRs, with three of them being based on GA. Three different rooms were employed, all of them being more reverberant than in the current study ($RT20(1kHz)>1s$ vs. $RT60(1kHz)=0.5s$). The BRIRs were notably evaluated in terms of authenticity, comparing them with measured BRIRs via an A-B test. This particular study revealed significant differences between the simulated BRIRs and a reference based on measured BRIRs in terms of authenticity. These findings might potentially be linked to the results of the current study. The current perceptual evaluation framework is similar to the one employed by Brinkmann et al. \cite{brinkmann2019round}, involving a direct comparison among models in terms of authenticity with a reference through a MUSHRA paradigm. This study also hints that this paradigm possesses the ability to discern even subtle perceptual differences. In the Brinkmann et al. study, the authors demonstrate that the observed differences were caused by spectral differences (tone color) in the early and late reflections, and perceived source position. They also evaluated the plausibility of each simulated BRIR; three of the four tested models provided similar plausibility results when compared to the reference. The participants were exposed to a visual representation of the room on a television. These results also suggest that, even when generating variations compared to a reference, an auralisation has the potential to present itself as ``plausible'' within a specific room. It is important to note that ``plausibility'' differs from ``authenticity'' and ``similiarty''  which both imply being perceived exactly the same as a reference, without any noticeable differences.

\subsection{Effect of the band reduction}


The numerical analysis revealed that the frequency information related to absorption coefficients had a greater effect on objective measures than the geometry reduction. Figure \ref{fig:ASWLEV} shows how the LEV metric displays a clear trend toward being inversely related to the number of bands for positions 2 and 3. Similarly, the spectral divergence with the reference gradually increases when the number of bands per absorption coefficient decreases. Considering this criterion, it can be seen that 1-band and 2-band models are approximately identical. This was somewhat expected, as the frequency content of the reflections is linked to the frequency content of the absorption coefficients. Consistent with these numerical analyses, for positions 2 and 3, decreasing the number of bands led to a significant perceptual difference ($>$50 in mean ratings) even for the 4-band model. The observed difference is much larger than the one for GR models. This can be explained by looking at the stronger impact of the reduction of absorption coefficients on the presented numerical metrics. Moreover, in the reference model, the RT60 was fitted on 8 different frequency bands. This fitting was accomplished on fewer bands in the BR versions (4,2 and 1). As a result, the spectral profile of the reverberation time could not be precisely replicated in all BR models. Given the importance of recreating this spectral profile for auralisation, the lack of spectral similarity is most likely to be responsible for the larger differences in ratings observed between BR models and the references. It appears that decreasing the frequency information on absorption coefficients should be avoided when using GA for auralisation purposes, as its effect is much more obvious than geometry reduction, both from a numerical and a perceptual standpoint.

\subsection{Study limitations and future work}


The main goal of this study was to evaluate the perceptual impact of simplifications of a GA method for Ambisonics-based binaural reverberation rendering. The concrete objective of these simplifications is to run simulations that could require little information about the acoustic environment.

First, the results of this study relies on various positions and stimuli within a single room only. Moreover, the GA method employed here, involving ray tracing, is not the most appropriate method to use in real time and can hardly allow listeners movements, other than head orientations, within the simulated space due to its computational complexity. Nevertheless, the results from this study could be useful to inform real time processing approaches.

It would be interesting to test these results with beam-tracing algorithms, which can be envisioned as a derivative of the ray tracing approach employed here. These technologies have been shown to be better suited to real time auralisation and dynamic movements \cite{laine2009accelerated, poirier2017evertims}, but they must also deal with a trade-off between computational cost and the complexity of the geometry \cite{siltanen2008geometry}. Other technical approaches, such as the image-source method, could contribute to efficiently modeling early reflections within a reverberation module \cite{poirier2017evertims}. Finally, SDNs and hybrid approaches using image-source method and FDNs are two different methods for modeling reverberation at a minimal computing cost. In a FDN, the delays are typically chosen to be prime numbers, which helps to ensure that the network produces a dense, diffuse reverberation pattern. In contrast, SDNs use a combination of random and fixed delays, which allows them to model both the diffuse and specular components of sound reflections \cite{de2015efficient}. Both can be potentially extremely applicable to real-time use, and could benefit from the results of this study since they are highly optimised for shoebox room simulations. 


It is also important to consider that the perceptual ratings could be called into question. Employing a single scale in the MUSHRA rating allowed for direct comparisons between different models and a reference, regardless of whether the perceived difference in reverberation was tied to spectral or spatial properties. This study demonstrated that the reference was distinct from all GR models. Nonetheless, the significance of this difference in specific tasks may be called into doubt. It could be investigated whether the perceived difference can cause a room divergence effect \cite{werner2016summary} with the visual characteristics of the room. More broadly, and as mentioned above, it could be explored how this difference affects various tasks in an augmented reality scenario. As mentioned earlier, Brinkmann et al. \cite{brinkmann2019round} demonstrated that auralisations that are perceptually different could still attain a compelling plausibility for a given room. In the future, it will be interesting to explore the effects of these perceptual differences resulting from reverberation rendering simplifications. This can be conducted within more ecological VR scenarios. For example, it could involve a direct comparison of the simulation with real-world signals.

Moreover, the study relied on self-reported lack of hearing impairment without conducting objective hearing tests. This may introduce a bias, as self-reported data can be inaccurate or incomplete, as some participants might not be aware of mild or moderate hearing impairments.

Finally, looking at potential future advancements in this domain, perceptual auditory models could represent a very powerful tool for future research \cite{sondergaard2013auditory}.
The capacity of listeners to localise sounds in the context of anechoic binaural auralisation has already been successfully predicted using such models \cite{engel2022assessing}. However, models that account for reverberation are currently scarce and focus mostly on speech intelligibility tasks \cite{leclere2015speech}. Future studies would very much benefit from models that account for reverberation and focus on more subjective perceptual characteristics (e.g. externalisation, realism, etc.).

\section{\label{sec:ccl}Conclusions}

This paper looked at the impact of simplifying input parameters of a model used for the rendering of Ambisonics-based binaural reverberation generated using a GA approach. The study investigates two distinct simplification processes: geometry reduction and reduction of the frequency content of absorption coefficients. Across all models studied, the direct sound was replicated through convolution with a dense HRIR dataset, while the reverberation was encoded as a third-order Ambisonics sound field.

It was initially hypothesised that minor reductions in the room's geometry would not have a significant effect on the perception of its simulated reverberation. However, even the slightest reduction in geometry yielded a small but significant perceptual impact. Additionally, the results highlight a consistent difference in perceptual ratings between each reduced model (excluding the shoebox model) and the reference. This suggests that geometry reduction can be used to a certain extent without increasing the perceptual difference further. Moreover, the shoebox model was found as different to all other reduced models in terms of similarity ratings. It suggests that, in the specific case of this study, considering the shoebox as a simplification is inadequate, particularly when the distance from the source is significant. These differences can be attributed to the significant spectral differences in the RIRs generated by this specific model. The reduction in frequency content of absorption coefficients led to strong perceptual differences compared to the reference, demonstrating the need to avoid this approach and to maintain frequency-specific information in the modelling and rendering stages. Moreover, the differences between models were significant for the two longest source-listener distances, showing that the further away the listener and the source are, the more relevant and measurable the differences between models become. On a similar note, if rooms with a higher DRR were investigated, the findings might be generalised. 

Finally, these perceptual changes should be interpreted with caution as they are based on geometrical acoustics and Ambisonics-based RIRs only. Future research should look into the full implications of this perceptual discrepancy for real-time rendering methods based on a geometrical model and potentially including real RIRs measurements. This study's use of ray tracing can be perceived as an initial step preceding the exploration of other methods, such as SDNs or hybrid approaches that combine image-source methods and FDNs, especially those making a shoebox assumption of the geomtry of the room.

\section*{Acknowledgments}
This study was made possible by support from SONICOM (\url{www.sonicom.eu}), a project that has received funding from the European Union’s Horizon 2020 research and innovation program under grant agreement No. 101017743.

\section*{Author declarations}

The data that support the findings of this study are available from the
corresponding author upon reasonable request. The study and methods followed the tenets of the Declaration of Helsinki. Approval for the experiment design was given by the Science, Engineering, and Technology Research Ethics Committee (SETREC, reference: 21IC6923) of Imperial College London. Data acquisition followed the general data protection regulation. The authors confirm that there are no conflicts of interest in the presented work.

\bibliographystyle{IEEEtran}
\bibliography{IEEEabrv,preprintsample_resubmit}

\end{document}